# Mobile Database System: Role of Mobility on the Query Processing

*Samidha Dwivedi Sharma*[1]
[1]Department Of Computer Science & Application
Dr. H.S. Gour University
Sagar, MP (India)
samidhad2000@gmail.com

*D.r R.S.Kasana*[2]
[2]Department Of Computer Science & Application
Dr. H.S. Gour University
Sagar, MP (India)
irkasana7158@gmail.com

*Abstract*—The rapidly expanding technology of mobile communication will give mobile users capability of accessing information from anywhere and any time. The wireless technology has made it possible to achieve continuous connectivity in mobile environment. When the query is specified as continuous, the requesting mobile user can obtain continuously changing result. In order to provide accurate and timely outcome to requesting mobile user, the locations of moving object has to be closely monitored. The objective of paper is to discuss the problem related to the role of personal and terminal mobility and query processing in the mobile environment.

## I. INTRODUCTION

Recent advances in hardware technology and wireless communication networks have directed to the emergence of mobile database systems [1,2]. The mobile computing environment provides database applications with useful aspects of wireless technology, which is known as mobile databases. This advance technology has created a new age of nomadic database users. These users are accessing a database through a network. Basically, a user with a wireless connection to the information network does not require maintaining a fixed position in the network.

In mobile environment, elements of the network are very dynamic and can be extremely volatile. Consider a database representing information about moving objects and their position in addition to information about stationary objects [5]. For example, a mobile user looking for a restaurant will obtain different results based on the time and the place he/she issued the query. As the location of other devices changes with respect to other entities and data sources are constantly in movement it may not be possible to collect information about available data sources at any given point of time. As the mobility is the most distinguishing feature of the mobile computing system, location becomes an important piece of information for location-dependent queries [16, 17]. Query may be issued from a moving object (e.g., car of a mobile user) or from a stationary user. Consequently, the answer to a location dependent query may depend on the location of the mobile host (MH) which issued the query and/or the locations of the objects represented in the database. Therefore, an optimal query processing subsystem of a mobile database has to take the strategy used by the location management component into account for answering queries.

The remainder of the paper is organized as follows. In Section 2, we introduce architecture of mobile environment. In Section 3, we describe the role of mobility. Section 4 presents the effect of mobility in mobile environments. Section 5 we study the Query processing system in mobile environment. Finally section 6 concludes the paper.

## II. ARCHITEURE OF MOBILE ENVIRONMENT

Figure 1 shows the existing and widely architectural model of a system that supports mobile computing [19,20]. We have added a number of DBSs (database servers) to incorporate database processing of capability without affecting any aspect of the generic mobile network. A set of general purpose computers (PCs, workstations etc) are interconnected through a high speed wired network.

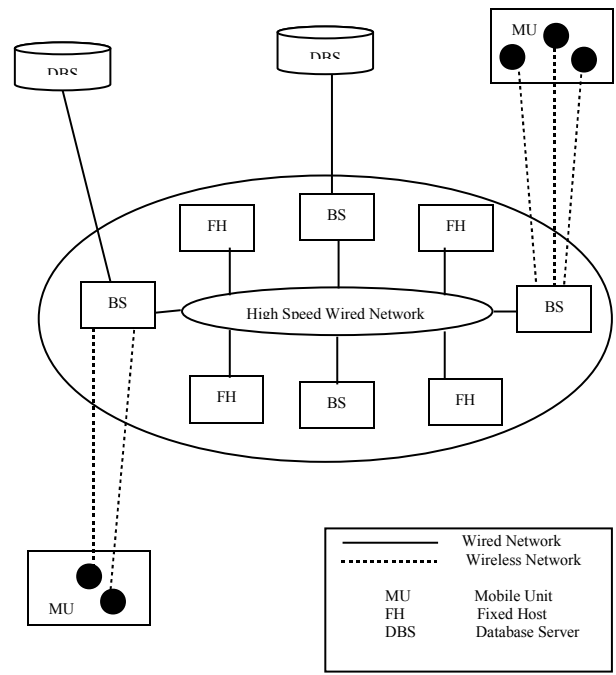

Figure1. Architecture of Mobile Environment

Components in the fixed network are either Fixed Hosts that are not capable of connecting to a mobile unit, or Base Stations which are computers capable of connecting with a mobile unit and are equipped with a database server and wireless interface (they are also known as Mobile Support Stations).A number of mobile computers (laptop, PDAs, etc) referred to as Mobile Hosts (MH) or Mobile units (MU) are connected to wired network components only through BSs via wireless channels. A BS maintains and communicates with its MUs via the wireless interface [4] and has some processing capability. Base Stationed hosts are connected together via a high-speed network (Mbps to Gbps).

### III. Role of Mobility

The location of mobile units is an important parameter when locating a mobile station that may hold the required data and when selecting information especially for location dependent information services [15, 18]. In this case the search cost, to locate mobile units, is added to the cost of each communication involving them. A mobile framework is composed of wired and wireless components and human users. The mobile discipline define two types of mobility :(i) terminal mobility and (ii)personal mobility.

*A. Terminal Mobility:*

In terminal mobility, the connection is established between two points and not between the two persons calling each other. This type of connection in a session allows the use of communication devices to be shared among anybody. It allows a mobile unit (laptop, cell phone, PDA, etc.) to access desired services from any location while in motion or stationary, irrespective of who is carrying the unit. For example a cell phone can be used by its owner and it can also be borrowed by some one else for use. In terminal mobility, it is the responsibility of the wireless network to identify the communication device.

Fig 2 illustrates the notion of terminal mobility. A person at location X (Longitude/Latitude = X) uses the mobile unit to communicate with the car driver at location P. He can still establish communication with the driver from a new location Y irrespective of the movement of the car from P to Q. The use of a phone card works on this principle. It can be used from different locations and from different machines such as cell phones, residential phones, etc.

*B. Personal Mobility*

In personal mobility the mobility of a person is supported. Thus, a user does not have to carry any communication equipment with him; he/she can use any communication device for establishing contact with the other party. This facility requires an identification scheme to verify the person wishing to communicate. Figure 3 illustrates the notion of personal mobility. P person at location X communicates with the car at location P using his PDA, and from location Y also he/she can communicate with the car at location P using his laptop. At present, personal mobility is available through the web. A user can log on to the web from different machines located at different places and access his/her e-mail. The mobile system extends this facility so that the user can use any mobile device for reaching the internet. In personal mobility each person has to be uniquely identified, and one way to do this is via a unique identification number. There is no dependency relationship between terminal and personal mobility; each can exist without the other. In personal mobility the party is free to move, and in terminal mobility the communication unit is free to move. Voice or data communication can be supported by either types of mobility. However, to visualize a complete mobile database management system both types of mobility are essential.

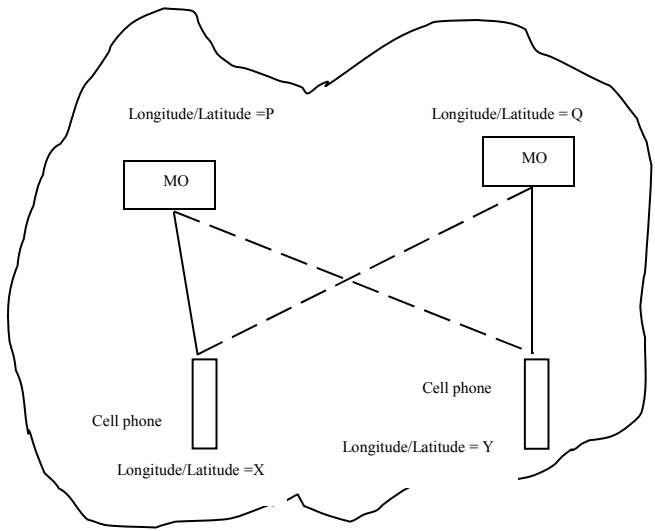

MO-Mobile Object (Car, Bike etc)

Figure 2. Terminal Mobility

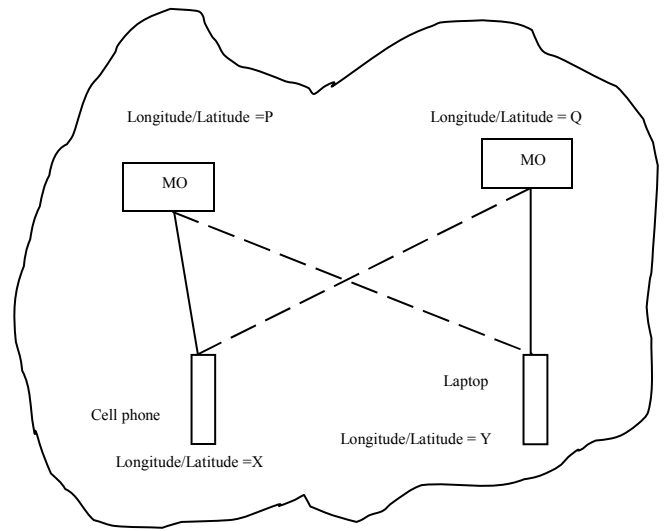

MO-Mobile Object (Car, Bike etc)

Figure 3. Personal Mobility

## IV. Effect of mobility on Query processing

The effects of mobility on query processing require that algorithms employed must be capable of managing frequent loss and appearance of mobile device in the network, and that overhead should be minimized during periods of low connectivity [7, 10]. In this environment we can distinguish many characteristics that are listed in the next sub sections.

### A. Different Location Models

Location-based applications require a well-formed representation of spatial knowledge. Current location models can be classified into symbolic or geometric models. In geometric model locations are specified as an n-dimensional coordinate (longitude-latitude pair) or a set of coordinates defining an area's bounding geometric shape (such as a polygon). Symbolic models refer to a location by some abstract symbols. Such a representation allows a reference to a place simply by abstract symbol or name, which makes it very convenient for human interaction [14].

### B. Different Query Types

In this section, we classify type of queries in mobile databases. The queries can be entirely new and specifically applied in the wireless environment, while the other can be a common type of query in traditional databases. The mobility in a mobile environment introduces three types of entities: (i) mobile client that submits a query, (ii) mobile server that processes a query or a part of it, and (iii) moving object which represents the data targeted by the query. According to these entities queries can be classified into five categories.

*1) Non Location Related Query (NLRQ)*

If all the predicates and attributes used in a query are non location related then it is called a Non Location Related Query (NLRQ). For example: "select all restaurants with South India specialty".

*2) Location Dependent Queries (LDQ)*

If a query results depend on the location of the query issuer then the query is called Location Dependent Query [8]. For example: "Find me the closest hotel within 10 miles of my current position".

*3) Location Aware Query (LAQ)*

If a query has at least one Location Related simple predicate or one Location Related attribute then it is called Location Aware Query[3]. For example: "How is the weather in Hyderabad?".

*4) Continuous Query (CQ)*

This type of queries includes all queries issued by mobile terminals and querying objects which are themselves moving. For example a query of this type could be : "Find all the cars within 100 feet of my car". The result of the query is a set of cars position that varies continuously with the movement of the driver.

*5) Ad Hoc Query*

Ad Hoc queries are commonly utilized queries in traditional DBMS. This type of query explicitly mentions the required information in the query statement, and does not involve any context awareness information. Thus, the query result is only based on the actual query itself. For example a query of this type could be: "University student wants to retrieve his/her academic record or personal details.

### C. Query Optimization

Query optimization methods try in general to obtain execution plans which minimize CPU, input/output and communication costs. In centralized environments the cost that affects most is the input/output whereas in distributed environments, communication cost is the most important. In a mobile distributed environment, the communication costs are much more difficult to estimate because the mobile host may be situated in different locations. The best site from which to access data depends on where the mobile computer is located. In general, it is not worth calculating plans and their associated costs statically, but rather, dynamic optimization strategies are required in this mobile distributed context.

### D. Query Execution

In static systems, query processing execution sites are determined in advance, i.e., which steps are performed on the client and which one on the server. In a mobile environment, where users are moving, such assumption is inadequate. Thus, mobile database systems must be able to choose an execution site for the different phases of query processing depending on their current environment and should be able to revise that decision as flexible as possible.

### E. The Impact of Portable Devices Limitations

If we reference dynamic location information in a query, we have to use a location management component to get this information. Thus, depending on the offered localization strategy, we have different possibilities to use this information. The cost evaluation of a query execution plan is guided by required resources of the plan. The main factors that are used in stationary systems are CPU-usage and the number of hard disk access. In mobile systems, additional varying factors like energy consumption, available memory and CPU-speed may be included.

### F. The Impact of Wireless Communication

The new networking technologies allow spontaneous connectivity among mobile devices, including hand helds, computers in vehicles, computers embedded in the physical infrastructure, and (nano) sensors. Mobile devices can suddenly become both sources and consumers of information [13]. There is no longer a clean distinction between clients and servers, instead devices are now peers. Furthermore, there is no longer a guarantee of infrastructure support. Consequently, for obtaining data, devices cannot simply depend on a help of some fixed, centralized server. Instead, the devices must be able to cooperate with others in their proximity in order to pursue individual and collective tasks.

## V. QUERY PROCESSING IN MOBILE DATABASES

### A. Current Work on Query Processing

Query processing deals with designing algorithms that analyze queries and convert them into a series of data manipulation operations. The main function of a query processor is to transform high-level query (typically, in relational calculus) into an equivalent lower-level query (typically, in some variation of relational algebra). Figure 3 shows the classic architecture for query processing. This architecture can be used for any kind of database system including centralized, distributed, or parallel systems. The query processor receives a query as input, translates and optimizes this query in several phases into an executable query plan, and executes the plan in order to obtain the results of the query.

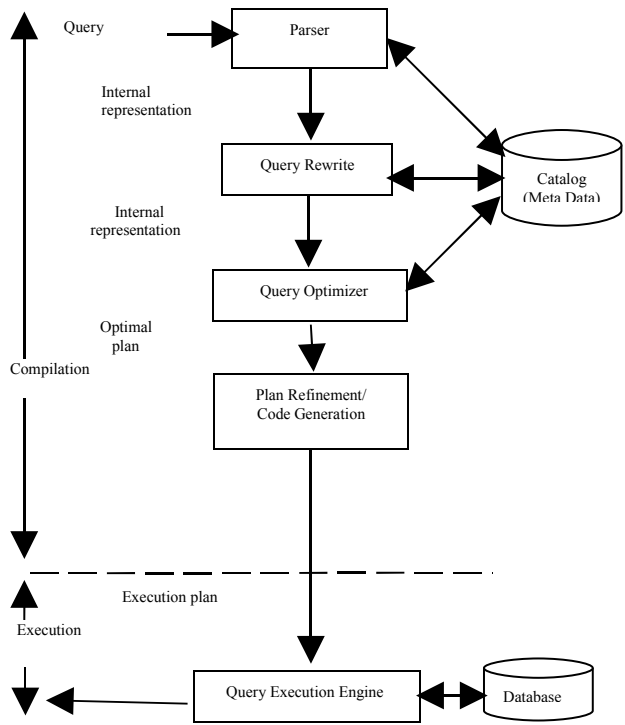

Figure 4. Phase of Query Processing

Not much work has been done and reported on query processing in mobile environment so for. But most of then maintain an approximate information about MUs location which increases the paging cost. MUs are able to access locally unavailable files/data from other remote BSs at the cost of huge message exchange [6, 9, 12].

### B. Types of Query Processing Systems

The query processing systems are able to give answer of different types of query by exchanging a less number of messages among the various component of the network. Figure 5 shows the query processing in mobile databases.

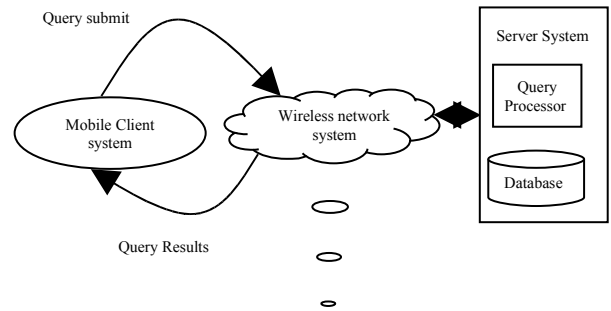

Figure 5. Query Processing in Mobile databases

We divide query-processing system for mobile databases into three parts, namely: (i) mobile client system, (ii) wireless network system and, (iii) server system.

Mobile clients system relates to how client manipulates and maintains the data in its cache efficiently and effectively. Wireless network system communicates data using broadcasting systems [11]. With this system, the number of mobile users does not affect the query performance. Server system relates to designing techniques for the server to accommodate multiple requests so that the request can be processed as efficiently as possible. We say that query processing for mobile databases is very much centered around the issues of caching, broadcasting, and scheduling.

*1) Mobile Client System*

Each mobile client is composed of three modules as shown in figure 6 a Resource Manager which manage the client CPU for handling the query results, a Client Manager which processes the query requests and passes them to the server, models the disconnection operation, and receives and processes the tuples transmitted from the server and a Query Generator which generates the query requests, and Client queries are submitted from an MH to the server to be processed and a message (messages) containing the tuples that form the answer to the query is (are) transmitted back to the MH. The messages containing the tuples are processed by the MH and the tuples are displayed on the screen of the MH accordingly.

Mobile client system defines a number of strategies to maintain cached data items in client's local storage. Basically, wireless communication channel in general suffers to narrow bandwidth while it is also inherently asymmetric communication, in which the downstream communication bandwidth from a server to the mobile client is much larger than the upstream communication bandwidth from clients back to server. Due to the above reason, caching of frequently accessed data items in a client's local storage becomes important for enhancing the performance and data availability of data access queries. Another advantage of caching is the ability to handle fault tolerance. This is related to the characteristics of mobile computing in which each mobile client connects or disconnects from the network frequently. In some situation, the BS may not be accessible due to problem like signal distortion. However, a query can still be partially

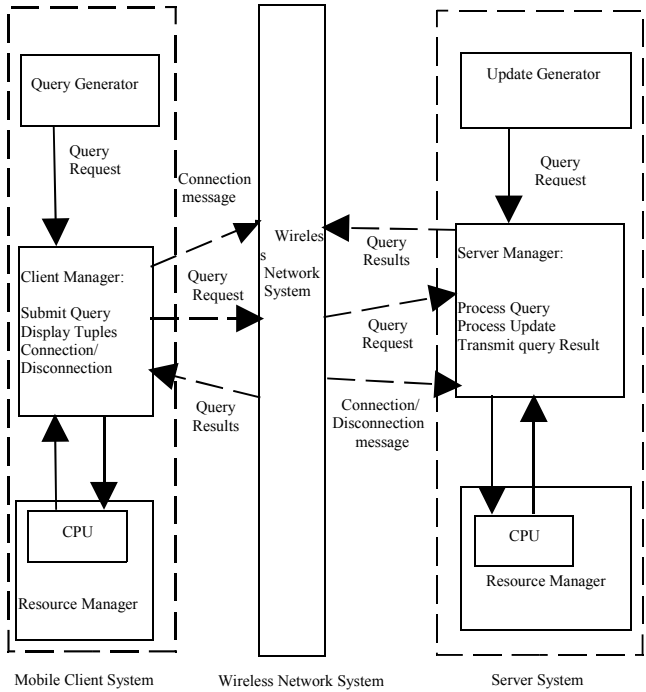

Figure 6. Complete System for Mobile Query Processing

processed from caches and at least some of the query results can be returned to the user.

*2) Wireless network System*

The Wireless Network System component assumes that all messages are of equal priority that will be served on a First-Come First-Served (FCFS) basis with a service rate of Network Bandwidth. When the Wireless Network Manager finds out (i.e., while sending a message to an MH) that an MH is disconnected, it informs the Server Manager about the disconnection so that the transmission of the tuples to the MH can be paused until the MH reconnects to the network. In this scheme, the server also periodically broadcasts the frequently accessed data items to clients through one or more broadcast channels, but the clients may not cache the data item of interest. This situation might occur when the client does not have enough memory or space to cache the required data.

*3) Server System*

This system considers the problem of pull-based broadcast scheduling where mobile clients send queries to the server through wireless network system, server process the query, and send the result back to the client. The server system has three subsystems as shown in Fig. 6. A Resource Manager which manage the server CPU time for query and update processing, an Update Generator which generates update requests, and a Server Manager which coordinates the query requests from MHs and update requests from the Update Generator.

When a CQ is issued by an MH, it is processed by the Server Manager and the set of tuples satisfying the query are determined. The number of tuples in the answer set of a CQ is determined randomly using a maximum value of Maximum Number of tuples. The Server Manager also decides when and which tuples should be transmitted. The strategy concerns with broadcast and disk scheduling. Broadcast scheduling is to determine how queries to be served efficiently in the server considering a number of factors such as the length of the query, the wait time and the popularity of the items.

A database server is able to enhance the data retrieval performance by incorporating its own main memory and cache to store those database items frequently accessed by most mobile clients. A query can be processed either in the disk server or cache server. If the relevant data items have been retrieved earlier then the query is processed in the cache server. After processing a query, the results are transmitted to the transmitter queue, which subsequently send the data items through the wireless channel.

## VI. CONCLUSION

Recent developments in wireless technology enables nomadic people are now able to access email, news, weather, and query to the central database server using wireless devices. The present work gives response to different types of queries with the help of mobile database stored at BS. Mobile database focuses on the query issue that is the main operation in mobile computing. Since mobile database is a new dimension of database application, the types of query, query processing system, and communication technology that involves in the application are different than what applies in traditional databases. We have defined role of mobility as well as query processing system in mobile databases. Query processing in mobile databases includes mobile client, wireless network, and server system.

The present work may be extended in future by capturing the query arrival pattern in the various BSs of the environment, to minimize the number of message exchanges and delay in getting answer for such query. A secure online mobile database and optimized query processing system needs to be developed to understand this possible as a reality.

## AUTHORS PROFILE

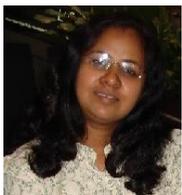
**Samidha Dwivedi Sharma** is currently a research scholar at Department of Computer Science and Applications, Dr. H. S. Gour Central University (formerly, Sagar University) Sagar, M P, India. She completed her bachelor's degree in Science (B.Sc.) with Mathematics subject in 1992 from Dr. H. S. Gour Central University (formerly, Sagar University) Sagar, M P, India. She received her Master degree in Computer Application in the year 1997 from Barkatullah University, Bhopal, India. Her fields of interests are Database Management Systems, Mobile database, Data Structure and mobile computing. She has published more than 4 research papers. She is a life member of ISTE.

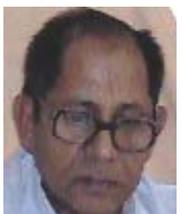
**Dr. R. S. Kasana** completed his bachelor's degree in 1969 from Meerut University, Meerut, UP, India. He completed his master's degree in Science (M.Sc.-Physics) and master's degree in technology (M. Tech.-Applied Optics) from I.I.T. New Delhi, India. He completed his dorctoral and post doctoral studies from Ujjain University in 1976 in Physics and from P. T. B. Braunschweig and Berlin, Germany & R.D. Univ. Jabalpur correspondingly. He is a senior Professor and HoD of Computer Science and Applications Department of Dr. H. S. Gour University, Sagar, M P, India. During his tenure he has worked as vice chancellor, Dean of Science Faculty, Chairman Board of studies. He has more than 34 years of experience in the field of academics and research. Twelve Ph. D. has awarded under his supervision and more than 110 research articles/ papers have published.